# Dynamic Spin Injection into Chemical Vapor Deposited Graphene

A. K. Patra[1,a)], S. Singh[2,a)], B. Barin[2], Y. Lee[3], J.-H. Ahn[3], E. del Barco[2,b)], E. R. Mucciolo[2] and B. Özyilmaz[1,4,5,6,b)]

[1]Department of Physics, National University of Singapore, 2 Science Drive 3, Singapore 117542

[2]Department of Physics, University of Central Florida, Orlando, Florida USA, 32816

[3] School of Advanced Materials Science & Engineering, SKKU Advanced Institute of Nanotechnology (SAINT), Sungkyunkwan University, Suwon, Republic Korea 440746

[4]NanoCore, 4 Engineering Drive 3, National University of Singapore, Singapore 117576

[5]Graphene Research Center, National University of Singapore, Singapore 117542

[6]NUS Graduate School for Integrative Sciences and Engineering (NGS), National University of Singapore, Singapore 117456

We demonstrate dynamic spin injection into chemical vapor deposition (CVD) grown graphene by spin pumping from permalloy (Py) layers. Ferromagnetic resonance measurements at room temperature reveal a strong enhancement of the Gilbert damping at the Py/graphene interface, exceeding that observed in even Py/platinum interfaces. Similar results are also shown on Co/graphene layers. This enhancement in the Gilbert damping is understood as the consequence of spin pumping at the interface driven by magnetization dynamics. Our observations suggest a strong enhancement of spin-orbit coupling in CVD graphene, in agreement with earlier spin valve measurements.

[a)]A. K. Patra and S. Singh contributed equally to this work.

[b)] Authors to whom correspondence should be addressed. Electronic mails: delbarco@physics.ucf.edu and phyob@nus.edu.sg

In spintronics, where the electron's spin degree of freedom, rather than its charge, is employed to process information, the efficient generation of the large spin currents stands as a key requirement for future spintronic devices and applications. Several approaches to generate pure spin currents have been proposed and are being widely investigated, namely, non-local spin injection [1], spin Hall effect [2-4], and spin pumping [5,6]. Among these, spin pumping offers the advantage of producing spin currents over large (mesoscopic) areas [7-13] at ferromagnetic/non-magnetic (FM/NM) interfaces. In addition, dynamical spin pumping is insensitive to a potential impedance mismatch at the FM/NM interface [14], a problem ubiquitous in the non-local spin injection approach. Dynamical spin pumping consists of generating pure spin current (i.e., with no net charge current) away from a ferromagnet into a non-magnetic material, induced by the coherent precession of the magnetization upon application of microwave stimuli of frequency matching the ferromagnetic resonance (FMR) of the system [15]. Since pure spin currents carry away spin angular momentum, in an FMR experiment the transfer of angular momentum from the FM into the NM layer results in an enhancement of the Gilbert damping in the ferromagnet [5-15]. Most studies of dynamical spin pumping on FM/NM interfaces have made use of Pt and Pd NM layers, since the large spin-orbit coupling in these systems enables the conversion of the injected spin current into an electric voltage across the NM layer, a phenomenon known as inverse spin Hall effect (ISHE). Recently, spin pumping has been experimentally demonstrated in FM/semiconductor interfaces (e.g., GaAs [13] and p-type Si [15]). However, there is no experimental report on spin pumping in FM/graphene interfaces, though graphene [16] (a two-dimensional layer of carbon atoms), possesses unique electronic properties (e.g. high mobility and gate-tunable charge carrier, among others), and stands as an excellent material for spin transport due to its large spin coherence length [17].

In this Letter we report experimental FMR studies of Py and Co films and polycrystalline graphene grown by chemical vapor deposition on Cu foils [18,19] (henceforth, Co/Gr and Py/Gr, respectively) performed in a broad-band microwave coplanar waveguide (CPW) spectrometer. The observation of a remarkable broadening of the FMR absorption peaks in the Py/Gr (88%) and Co/Gr (133%) films demonstrate a strong increase of the Gilbert damping in the FM layer due to spin pumping at the FM/Gr interface and the consequent loss of angular momentum through spin injection into the CVD graphene layer. To account for such a remarkable absorption of angular momentum, the spin orbit coupling in CVD graphene must be orders of magnitude larger than what is predicted for pristine, exfoliated graphene.

To prepare the FM/Gr samples, single layer CVD grown graphene [18, 19] was first transferred onto a Si substrate with 300 nm thick $SiO_2$ layer. The sample was then annealed in a $H_2$/Ar environment at 300°C for 3 hour to remove all organic residues. For the Py layer we chose $Ni_{80}Fe_{20}$, a material extensively used for magnetic thin film studies because of its low magnetocrystalline anisotropy and its insensitivity to strain. The FM layer (Py/14nm, Co/15nm) was deposited on top of the graphene layer lying over the $SiO_2$/Si substrate by electron-beam evaporation at a base pressure of $3\times10^{-7}$ Torr. For the purpose of FMR comparison experiments, a control FM film of the same thickness was deposited simultaneously on the same $SiO_2$ wafer in an area where graphene was not present. The schematic of the FM/Gr samples is shown in Fig. 1-a, together with the Raman spectrum of the CVD graphene before the deposition of Py (Fig. 1-c). The high intensity of the 2D peak, when compared to the G peak, and the weakness of the D peak, suggests that graphene is single layer and of high quality (i.e. low degree of inhomogeneity/defects).

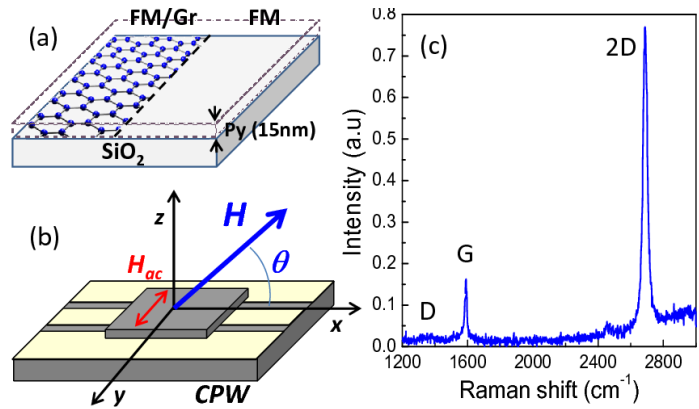

**Fig. 1**: (Color online) (a) Schematic of the FM/Gr film sample. (b) Schematic of the FMR measurement setup, with the sample placed up-side-down on top of the micro-CPW. (c) Raman spectrum of CVD graphene.

FMR measurements were carried out at room temperature with a high-frequency broadband (1-50 GHz) micro-coplanar-waveguide (µ-CPW) [20] using the flip-chip method [21-23], by which the sample is placed up-side-down covering the central part of the CPW (as shown in Fig. 1-b), where the transmission line is constricted to increase the density of the microwave field and enhance sensitivity. The CPW was covered with a 100nm-thick insulating layer of PMMA resist, hardened by electron beam exposure, to avoid any influence of the CPW, made out of gold, on the sample dynamics. A 1.5 Tesla rotatable electromagnet was employed to vary the applied field direction from the in-plane ($\theta = 0°$) to normal-to-the film plane ($\theta = 90°$) directions. Fig. 2-a shows the angular dependence of the FMR field measured at 10 GHz for both Py and Py/Gr films. The rotation plane is chosen to keep the dc magnetic field, $H$, perpendicular to the microwave field felt by the sample at all times, as shown in Fig. 1-b. The resonance field increases as the magnetic field is directed away from the film plane (i.e. increasing $\theta$), as expected for a thin film ferromagnet with in-plane shape magneto-anisotropy. The angular dependence of the FMR field ($H_R$) can be fitted using the resonance frequency condition given by the Smit and Beljers formula [23,24],

$$\omega = \gamma\sqrt{H_1 H_2}, \qquad (1)$$

where $\omega = 2\pi f$ is the angular frequency, $\gamma = g\mu_B/\hbar$ the gyromagnetic ratio, and $H_1$ and $H_2$ are given by

$$H_1 = H\cos(\theta - \varphi) - 4\pi M_{eff}\sin^2\varphi$$
$$H_2 = H\cos(\theta - \varphi) + 4\pi M_{eff}\cos 2\varphi + \frac{2K_2}{M_S}\sin^2\varphi, \qquad (2)$$

where $\varphi$ is the magnetization angle, $4\pi M_{eff} = 4\pi M_S - 2K_1/M_S - 4K_2/M_S \cos^2\varphi$ is the effective demagnetization field, $M_s$ is the saturation magnetization, and $K_1$ and $K_2$ are the first and second order anisotropy energies, respectively. The best fits to the data in Fig. 2-a are given by the parameters shown in the third column of Table 1, together with the corresponding parameters extracted from equivalent measurements on the Co and Co/Gr films (not shown).

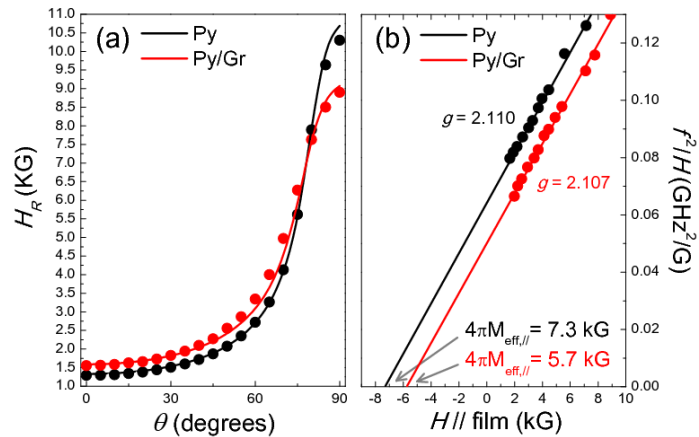

**Fig. 2**: (color online) (*a*) Angular dependence of the FMR fields measured on both Py and Py/Gr samples at $f$ = 10 GHz with the dc magnetic field, $H$, applied in a plane perpendicular to the microwave field generated by the CPW at the sample position. (*b*) In-plane frequency dependence of the FMR fields for both Py and Py/Gr samples. The intercepts with the *x*-axis give the effective demagnetizing fields of the samples.

It is useful to study the resonant behavior by applying the magnetic field at $\theta = 0°$ (parallel configuration) and $\theta = 90°$ (perpendicular configuration), since the frequency behavior of the FMR fields are given respectively by,

$$\left(\frac{\omega}{\gamma}\right)^2_{//} = H_R(H_R + 4\pi M_{eff,//})$$
$$\left(\frac{\omega}{\gamma}\right)_{\perp} = H_R - 4\pi M_{eff,\perp} \quad , \quad (3)$$

where $\gamma = g\mu_B/\hbar$ is the gyromagnetic ratio, $4\pi M_{eff,//} = 4\pi M_S + H_{A1}$, $4\pi M_{eff,\perp} = 4\pi M_S + H_{A1} + H_{A2}$, with $H_{A1} = 2K_1/M_S$ and $H_{A2} = 4K_2/M_S$ the first and second order anisotropy fields, respectively, which relate to surface, interface and/or magnetoelastic anisotropy. Note that $K_1 > 0$ ($\gg K_2$) provides out-of-plane anisotropy, competing with the in-plane shape anisotropy. Consequently, a graphical representation of the in- and out-of-plane frequency response of the FMR fields, conforming to Eq. (3), results in a linear behavior from which the slope and intercept with the magnetic field axis give $\gamma$ and the effective demagnetization fields, respectively. The results obtained for the Py and Py/Gr samples are shown in Fig. 2-b and 2-c, and the extracted parameters are listed in the third column of Table 1, together with those extracted from the Co and Co/Gr. Note that the anisotropy fields depend on the selection of the saturation magnetization, with theoretical values $M_{S,Py} = 9.27$ kG (attending to a 20/80-Ni/Fe ratio and assuming identical densities), and $M_{S,Co} = 17.59$ kG. For the Co and Co/Gr films, the effective saturation magnetization ($M_{eff} = 17.7$ kG) is similar to the one expected from theory, hence there is negligible out-of-plane anisotropy ($K_1 \sim 0$), in agreement with previous studies [25]. The situation is different in the case of the Py and Py/Gr, where the small Py anisotropy field $H_{A1} = 1.98$ kG grows significantly in the Py/Gr ($H_{A1} = 3.60$ kG), suggesting an increase of the Py surface anisotropy due to the presence of the graphene layer (i.e.

interface effect). Nevertheless, the magnetization remains in the plane of the film for all samples.

| Theory | Sample | $H_R$ vs. $\theta, f$ | Damping | Changes |
|---|---|---|---|---|
| **Py: $Ni_{80}Fe_{20}$** $g_{eff} = 2.10$ $g_{eff} = \dfrac{0.8 M_S^{Ni} + 0.2 M_S^{Fe}}{0.8 M_S^{Ni}/g_{Ni} + 0.2 M_S^{Fe}/g_{Fe}}$ $M_s = 9.27$ kG $Ms = 0.8 M_S^{Ni} + 0.2 M_S^{Fe}$ with $M_S^{Ni} = 6.094$ kG  $g_{Ni} = 2.21$ $M_S^{Fe} = 22.016$ kG  $g_{Fe} = 2.0$ | Py | $g = 2.110$ | $\alpha = 0.0113$ $G = 0.311$ GHz | $K_1$ increases (interface) Damping increases by ~88% |
| | | $M_{eff} = 7.30$ kG | | |
| | | $H_1 = 1.98$ kG | | |
| | | $K_1 = 0.73 \times 10^6$ erg/cc | | |
| | Py/Gr | $g = 2.107$ | $\alpha = 0.0213$ $G = 0.585$ GHz | |
| | | $M_{eff,//} = 5.70$ kG | | |
| | | $H_{A1} = 3.60$ kG | | |
| | | $K_1 = 1.32 \times 10^6$ erg/cc | | |
| **Co** $g = 2.145$ $M_s = 17.59$ kG | Co | $g = 2.149$ | $\alpha = 0.0210$ $G = 1.11$ GHz | (no $K_1$) Damping increases by ~133% |
| | | $M_{eff} = 17.7$ kG | | |
| | Co/Gr | $g = 2.149$ | $\alpha = 0.0489$ $G = 2.59$ GHz | |
| | | $M_{eff} = 17.5$ kG | | |

**TABLE I**: Parameters extracted from the analysis of the data reported in this work.

We now focus on the FMR linewidth and its frequency dependence when the magnetic field is applied parallel to the film ($\theta = 0°$), from which information about the Gilbert damping (i.e., spin relaxation dynamics) can be directly extracted. The inset to Fig. 3 shows a field derivate of the CPW $S_{21}$ transmission parameter obtained when exciting the FMR at 10 GHz in both Py and Py/Gr samples, with $H_R = 1.28$ kG and 1.55 kG, respectively. The peak-to-peak distance represents the linewidth, $\Delta H$, of the FMR, whose behavior as a function of frequency is shown for both samples in the main panel of Fig. 3. A remarkable increase of the FMR linewidth by 88% is observed in the Py/Gr sample, and even higher (133%) in the Co/Gr films. The change in the linewidth must be attributed to a substantial enhancement of the Gilbert damping in the FM film due to the influence of the graphene directly underneath. The frequency dependence of the FMR linewidth can be written as a contribution from two parts:

$$\Delta H = \Delta H_0 + \frac{4\pi\alpha}{\sqrt{3}\gamma} f, \qquad (4)$$

where $\alpha$ is the parameter of the Gilbert damping $G = \alpha\gamma M_S$. The first term, $\Delta H_0$, accounts for sample-dependent inhomogeneous broadening of the linewidth and is independent of frequency, while the second term represents the dynamical broadening of the FMR and scales linearly with frequency.

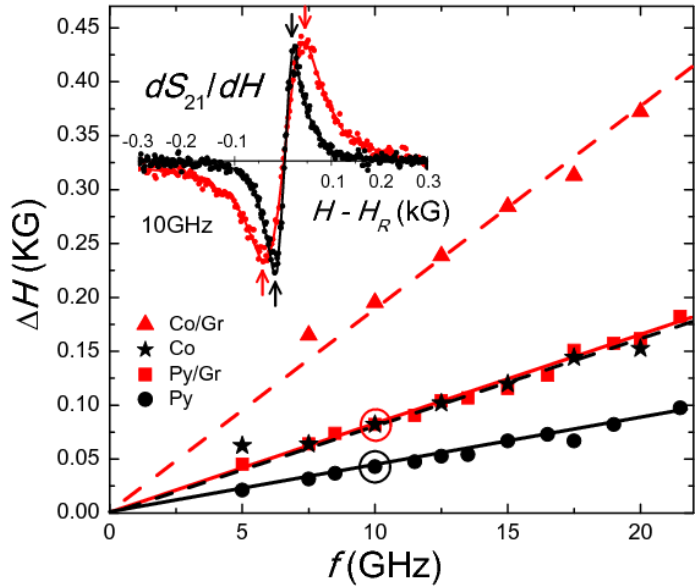

**Fig. 3**: (color online) Frequency dependence of the FRM linewidth for Py, Py/Gr, Co and Co/Gr films obtained with the magnetic field applied at $\theta = 0$ (in-plane configuration). The inset shows the field derivatives of CPW $S_{21}$ transmission parameter (at 10 GHz) of the Py and Py/Gr samples, from which the linewidth, $\Delta H$, is calculated as the peak-to-peak distance.

As observed in Fig. 3, the measured linewidth for both FM and FM/Gr samples increases linearly with frequency, with negligible inhomogeneous broadening, indicating that damping in the FM film can be properly explained by the phenomenological Landau-Lifshitz-Gilbert damping model. A similar broadening of the FMR linewidth is observed in both samples when the field is applied perpendicular to the plane, excluding frequency-dependence inhomogeneous broadening (e.g. two-magnon scattering produced by changes in morphology of the FM surface [26]), as its possible source. By fitting the data in Fig. 3 to Eq. (4) (using $\Delta H_0 = 0$), the damping parameters α and G are determined and given in the fourth column of Table 1 for all studied samples. The Gilbert damping increases substantially in the FM/Gr films as a result of the increased linewidth, when compared to the values obtained in the FM samples (which are comparable with values given in the literature for similar Py and Co films [9,22]). This is our key finding. Remarkably, the change in the damping parameter in the Py/Gr sample ($\Delta\alpha = \alpha_{Py/Gr} - \alpha_{Py} = 0.01$) is even more pronounced than those observed in Py/Pt systems, in which the thick (when compared to graphene) heavy transition metal Pt layer provides the large spin-orbit coupling necessary to absorb (i.e., relax) the spin accumulation pumped away from the ferromagnet. The efficiency of spin injection is usually cataloged by means of the interfacial spin-mixing conductance, which is proportional to the additional damping parameter, Δα, as follows:

$$g_{\uparrow\downarrow} = \frac{4\pi M_S d_{FM}}{\gamma \hbar} \Delta\alpha, \qquad (5)$$

giving $g_{\uparrow\downarrow} = 5.26 \times 10^{19}$ m$^{-2}$ for our Py/Gr sample with the thickness of the Py film $d_{FM}$ = 14 nm. The Py/Gr value is substantially larger than those found in other Py/NM systems with a metallic

NM layer, e.g., $g_{\uparrow\downarrow} = 2.19\times10^{19}$ m$^{-2}$ in Py(Ni$_{81}$Fe$_{19}$:10nm)/Pt(10nm) [9] or $g_{\uparrow\downarrow} = 2.1\times10^{19}$ m$^{-2}$ in Py(Ni$_{80}$Fe$_{20}$:15nm)/Pt(15nm) [11]. Note that in the cited experiments, the spin-diffusion length of the non-magnetic layer (~10 nm for Pt) is smaller than the layer thickness. This is significant since it explains how the Pt layer is capable of dissipating the spin accumulation generated by the dynamical spin pumping, and account for the loss of angular momentum in the Py. In the case of graphene, the enhancement of the damping parameter is more complicated to understand. In a standard FM/NM metallic system, the spin current injected into the NM layer decays mainly perpendicularly to the interface [27], causing the enhancement of the damping parameter to depend on the ratio between the layer thickness and the spin-diffusion length in the NM. However, graphene has effectively zero thickness and, at least theoretically, a very weak intrinsic spin-orbit coupling. Therefore, the spin current must decay in a FM/Gr film parallel and not perpendicular to the interface. Furthermore, some sizable spin-orbit coupling must exist in CVD graphene films. The latter may also explain the generally observed very short spin relaxation times in lateral CVD graphene spin valves [28]. Recently, small levels of hydrogen [29] and copper adatoms [30] have been predicted to lead to a strong enhancement of the spin-orbit coupling, bringing it into meV range. Cu adatoms are certainly likely to be present in the CVD samples utilized in our experiments, pointing at a possible explanation for the large spin pumping effect observed in our FM/Gr films.

**References:**

[1] F. J. Jadema, A. T. Filip, and B. J. van Wees: *Nature* (London) **410**, 345 (2001)

[2] 2.T. Kimura, Y. Otani, T. Sato, S. Takahashi, and S. Maekawa, *Phys. Rev. Lett.* **98**, 156601 (2007).

[3] S. O. Valenzuela and M. Tinkham, *Nature* (London) **442**, 176 (2006).

[4] T. Seki, Y. Hasegawa, S. Mitani, S. Takahashi, H. Imamura, S. Maekawa, J. Nitta, and K. Takanashi, *Nature Mater*. **7**, 125 (2008).

[5] E. Saitoh, M. Ueda, H. Miyajima, and G. Tatara, *Appl. Phys. Lett*. **88**, 182509 (2006).

[6] O. Mosendz, J. E. Pearson, F. Y. Fradin, G. E. W. Bauer, S. D. Bader, and A. Hoffmann, *Phys. Rev. Lett.* **104**, 046601 (2010).

[7] S. Mizukami, Y. Ando, and T. Miyazaki, *Phys. Rev. B* **66**, 104413 (2002).

[8] B. Heinrich, Y. Tserkovnyak, G. Woltersdorf, A. Brataas, R. Urban, and G. E. W. Bauer, *Phys. Rev. Lett.* **90**, 187601 (2003).

[9] K. Ando, T. Yoshino, and E. Saitoh, *Appl. Phys. Lett.* **94**, 152509 (2009).

[10] Y. Kajiwara, K. Harii, S. Takahashi, J. Ohe, K. Uchida, M. Mizuguchi, H. Umezawa, H. Kawai, K. Ando, K. Takanashi, S. Maekawa, and E. Saitoh, *Nature* (London) **464**, 262 (2010).

[11] O. Mosendz, V. Vlaminck, J. E. Pearson, F. Y. Fradin, G. E. W. Bauer, S. D. Bader, and A. Hoffmann, *Phys. Rev. B*. **82**, 214403 (2011).

[12] K. Ando, Y. Kajiwara, S. Takahashi, S. Maekawa, K. Takemoto, M. Takatsu, and E. Saitoh, *Phys. Rev. B* **78**, 014413 (2008).

[13] K. Ando, S. Takahashi, J. leda, H. Kurebayashi, T. Trypiniotis, C.H.W. Barnes, S.Maekawa and E. Saitoh *et al. Nature Mater.* **10,** 655–659 (2011).

[14] Arne Brataas, Yaroslav Tserkovnyak, Gerrit E. W. Bauer, Paul J. Kelly, http://arxiv.org/abs/1108.0385.

[15] Kazuya Ando and Eiji Saitoh, *Nature Communications* **3**, 629 (2012).

[16] K. S. Novoselov, A. K. Geim, S. V. Morozov, D. Jiang, Y. Zhang, S. V. Dubonos, I. V. Grigorieva, A. A. Firsov *Science* **306**, 666 (2004).

[17] N. Tombros, C. Jozsa, M. Popinciuc, H. T. Jonkman, and B. J. van Wees, *Nature* **448** (7153), 571 (2007); T. Y. Yang, J. Balakrishnan, F. Volmer, A. Avsar, M. Jaiswal, J. Samm, S. Ali, A. Pachoud, M. Zeng, M. Popinciuc, G. Güntherodt, B. Beschoten, and B. Özyilmaz, *Phys. Rev. Lett.* **107** (4) (2011); Wei Han and R. Kawakami, *Phys. Rev. Lett.* **107** (4) (2011); B. Dlubak, M-B. Martin, C. Deranlot, B. Servet, S. Xavier, R. Mattana, M. Sprinkle, C. Berger, W. A. De Heer, F. Petroff, A. Anane, P. Seneor, and A. Fert, Nat. Phys. **8** (7), 557 (2012).

[18] X. Li, W. Cai, J. An, S. Kim, J. Nah, D. Yang, R. Piner, A. Velamakanni, I. Jung, E. Tutuc, S. K. Banerjee, L. Colombo, and R. S. Ruoff, *Science* **324** (5932), 1312 (2009).

[19] S. Bae, H. Kim, Y. Lee, X. Xu, J-S. Park, Y. Zheng, J. Balakrishnan, T. Lei, H. R. Kim, Y. I. Song, Y-J. Kim, K. S. Kim, B. Ozyilmaz, J-H Ahn, B. H. Hong, and S. Iijima, *Nat. Nano.* **5** (8), 574 (2010).

[20] W. Barry, I.E.E.E Trans. *Micr. Theor. Techn. MTT* **34**, 80 (1996).

[21] G. Counil, J. V. Kim, T. Devolder, C. Chappert, K. Shigeto and Y. Otani, *J. Appl. Phys.* **95**, 5646 (2004).

[22] J-M. L. Beaujour, W. Chen, K. Krycka, C. –C. Kao, J. Z. Sun and A. D. Kent, *Eur. Phys. J. B* **59**, 475-483 (2007).

[23] J. J. Gonzalez-Pons, J. J. Henderson, E. del Barco and B. Ozyilmaz, *J. Appl. Phys.* **78**, 012408 (2008).

[24] S. V. Vonsovskii, *Ferromagnetic Resonance*, Pergamon, Oxford, (1996).

[25] J-M. L. Beaujour, J. H. Lee, A. D. Kent, K. Krycka and C-C. Kao, *Phys. Rev. B* **74**, 214405 (2006).

[26] M. J. Hurben and C. E. Patton, *J. Appl. Phys*. **83**, 4344-4365 (1998).

[27] Y. Tserkovnyak, A. Brataas, G. E. W. Bauer, and B. I. Halperin, *Rev. Mod. Phys.* **77**, 1375 (2005).

[28] A. Avsar, T. Y. Yang, S. Bae, J. Balakrishnan, F. Volmer, M. Jaiswal, Z. Yi, S. R. Ali, G. Guntherodt, B. H. Hong, B. Beschoten, and B. Ozyilmaz, *Nano letters* **11** (6), 2363 (2011).

[29] A. H. Castro Neto and F. Guinea, *Phys. Rev. Lett.* **103**, 026804 (2009).

[30] C. Cao, M. Wu, J. Jiang, and H-P. Cheng, *Phys. Rev. B*. **81**, 205424 (2010).